# What if we had built a prediction model with a survival super learner instead of a Cox model 10 years ago?

Arthur Chatton,[1,*] Émilie Pilote,[1] Kevin Assob Feugo,[1] Héloïse Cardinal,[2,3,4] Robert W. Platt,[5,6,7] and Mireille E. Schnitzer[1,2,5]

[1] Faculté de Pharmacie, Université de Montréal, Montréal, QC, Canada
[2] Département de Médecine Sociale et Préventive, École de Santé Publique de l'Université de Montréal, Université de Montréal, Montréal, QC, Canada
[3] Département de Médecine, Faculté de Médecine, Université de Montréal, Montréal, QC, Canada
[4] Centre de recherche du Centre hospitalier de l'Université de Montréal, Montréal, QC, Canada
[5] Department of Epidemiology, Biostatistics and Occupational Health, McGill University, Montréal, QC, Canada
[6] Department of Pediatrics, McGill University, Montréal, QC, Canada
[7] Center for Clinical Epidemiology, Lady Davis Institute, Jewish General Hospital, Montréal, QC, Canada
[*] Corresponding author: Arthur Chatton, email: arthur.chatton@umontreal.ca

**Abstract**

Objective: This study sought to compare the drop in predictive performance over time according to the modeling approach (regression versus machine learning) used to build a kidney transplant failure prediction model with a time-to-event outcome.

Study Design and Setting: The *Kidney Transplant Failure Score* (KTFS) was used as a benchmark. We reused the data from which it was developed (DIVAT cohort, n=2,169) to build another prediction algorithm using a survival super learner combining (semi-)parametric and non-parametric methods. Performance in DIVAT was estimated for the two prediction models using internal validation. Then, the drop in predictive performance was evaluated in the same geographical population approximately ten years later (EKiTE cohort, n=2,329).

Results: In DIVAT, the super learner achieved better discrimination than the KTFS, with a tAUROC of 0.83 (0.79-0.87) compared to 0.76 (0.70-0.82). While the discrimination remained stable for the KTFS, it was not the case for the super learner, with a drop to 0.80 (0.76-0.83). Regarding calibration, the survival SL overestimated graft survival at development, while the KTFS underestimated graft survival ten years later. Brier score values were similar regardless of the approach and the timing.

Conclusion: The more flexible SL provided superior discrimination on the population used to fit it compared to a Cox model and similar discrimination when applied to a future dataset of the same population. Both methods are subject to calibration drift over time. However, weak calibration on the population used to develop the prediction model was correct only for the Cox model, and recalibration should be considered in the future to correct the calibration drift.

**Highlights**:
- Predictive performance dropped over time for both regression-based and machine-learning-based prediction models.
- The survival super learner achieved better discrimination at the cost of a poorer weak calibration than the Cox model on development data.
- Training the survival super learner with a loss function maximizing discrimination yielded the strongest miscalibration.
- More frequent recalibrations of the score are needed when using the survival super learner.

**Introduction**

The need for kidney replacement therapies is a major public health issue, with worldwide use expected to total 5,439 million patients (3,899 to 7,640 million) in 2030.[1] Kidney transplantation is considered the treatment of choice, improving both quality of life and life expectancy as compared to remaining on dialysis.[2,3] Still, there is a shortage of organs available for donation. Optimizing long-term post-transplantation care is crucial to limit the need for novel transplantations. Clinical prediction models (CPMs), or prognostic scores, can provide meaningful information for nephrologists and may help involve patients in managing their disease. When it comes to kidney transplantation, researchers often want to predict time-to-event outcomes like the return to dialysis (i.e., graft loss). For instance, Foucher et al. developed the Kidney Transplant Failure Score (KTFS) to predict the risk of graft failure seven years after the first anniversary of the transplantation,[4] which was recently externally validated.[5] Foucher et al. used a Cox model to build the KTFS, but alternative statistical methods can be used instead, such as accelerated failure time models,[6] random survival forests,[7] or deep learning.[8] All these methods will yield a different CPM, and researchers do not know *a priori* which one is the best for their application. The super learner (SL) is an averaging method combining predictions from various statistical methods (hereafter candidate learners) into one weighted prediction. It has been shown that the resulting prediction is asymptotically at least as accurate as the best candidate learners,[9] and was successfully applied in other domains.[10,11] However, it has seldom been used for time-to-event outcomes,[12,13] although it can handle such outcomes with right-censoring.[14,15]

Unfortunately, CPM performance comparisons focus almost solely on validity at the time of development.[12,16] However, populations naturally shift over time with, for instance, practice changes, such as the use of kidneys from donors who have risk features for suboptimal graft longevity.[17] Davis et al. studied the performance evolution of CPMs over time according to the statistical method of development for binary outcomes.[18] They showed that random forest and neural networks outperformed logistic regression (with or without penalization) in terms of calibration over time, while the discrimination remained similar. However, the censoring inherent to time-to-event outcomes may differentially impact the performance evolutions of regression and machine learning methods.

In this study, we sought to compare the evolution of the performance of a Cox-based CPM (the KTFS) to a survival SL-based CPM that includes random survival forests and survival neural networks. Both CPMs are built with the same data and validated temporarily on a similar population ten years later.

**Material and methods**

*Cox-based and survival SL-based CPMs*

The KTFS was developed on the French multicentric cohort DIVAT with 2169 patients included from 1997 to 2006 using a cause-specific proportional hazards Cox model.[4] The authors stated that yearly audits between centers reported less than 1% error in the collected data, minimizing the risk of measurement error. The study population was defined as adult recipients of neurologically deceased donors with a functional transplant on the first anniversary of their transplantation without missing predictors. The outcome was the time-to-return to dialysis over the following seven years, with censoring due to patient mortality. It includes as pre-transplant predictors recipient biological sex, recipient age, last donor creatininemia, and the number of previous transplantations; and four post-transplant predictors: creatininemia at 3 and 12 months, proteinuria at 12 months and acute rejection episode during the first year of transplantation. Note that this CPM is aimed to be used at the transplantation's first anniversary, explaining the post-transplant predictors. Non-linear functional forms or categorizations are used for continuous predictors, and interactions are also included.

We reused data from the same 2,169 individuals to build the survival SL-based CPM mimicking the KTFS. It included five regression methods – a Gamma-distributed AFT model, a Weibull-distributed AFT model, a Cox model with main terms only, an elasticnet Cox model,[19] and a Royston and Parmar flexible parametric proportional-hazards model[20] – in addition to the random survival forest and the survival neural network.[7,8] Therefore, the SL is based on methods varying in their model and distributional assumptions, and ways to incorporate continuous predictors and interactions, as recommended by Phillips et al.[21] The predictions from each method were weighted and combined with the objective of minimizing the Brier Score loss function at 7 years using a 10-fold cross-validation scheme. Hyperparameters were estimated using 10-fold cross-validation nested in each previous fold. The R package survivalSL was used to carry out the described procedures.[22] Package versions and values of the hyperparameters and seeds are provided in Supplementary Table 1.

*Performance evaluation of the CPMs*

The performances of the two CPMs were evaluated in DIVAT using bootstrap (2000 replications) for internal validation. Then, they were evaluated using the more recent French EKiTE data (transplantation between 2010 and 2015).[23] EKiTE is a European network that includes patients from France, Spain, Belgium, and Norway. We restricted our analysis to French patients to focus on the temporal evolution of performance without geographical disparity. We reused the eligibility criteria defined in the previous subsection and performed a complete case analysis. Therefore, we included 2329 individuals (Figure 1).

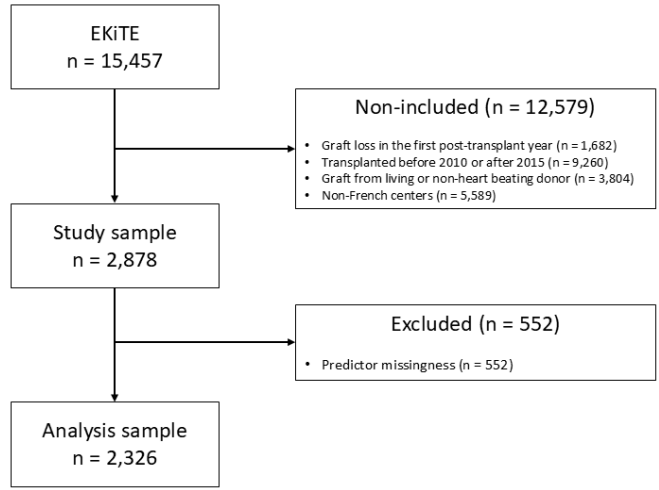

*Figure 1: Flowchart for the EKiTE network data analysis sample.*

The discrimination was evaluated by the area under the time-dependent receiver operating curve (tAUROC) at 7 years.[24] The calibration was evaluated at three different levels (mean calibration, weak calibration, and moderate calibration) as recommended by Van Calster et al.[25], and flexible calibration curves were fitted using restricted cubic splines with five knots. Overall performance was evaluated by the Brier Score. Right-censoring was considered random/uninformative, and inverse probability weighting was used to create a hypothetical population that would have been observed had no censoring occurred.[26] See McLernon et al. for more information on these metrics in survival settings.[27] Non-parametric bootstrap (2000 iterations) was used to compute 95%CIs.

*Sensitivity analyses*

We performed two sensitivity analyses. First, we allowed the survival SL to select the predictors using elasticnet[19] (which was removed from the pool of candidate learners) from the larger pool of predictors in DIVAT instead of using only the ones present in the KTFS. Second, we investigated two other loss functions in the survival SL instead of the Brier score: the AUROCt and the negative binomial log-likelihood.

**Results**

*Patients characteristics*

In DIVAT, the median follow-up time was 4.48 years, 182 (8.39%) patients returned to dialysis, and 1987 (91.61%) were right-censored. In EKiTE, the median follow-up time was 6.07 years, 242 (10.39%) patients returned to dialysis, and 2087 (89.61%) were right-censored. Survival was similar across the temporalities (Figure 2). As expected, patient characteristics differed between the two study cohorts (Table 1). For instance, more recent donors were older, with a mean age of 55.0 years (± 16.5) compared to 45.2 years (± 15.8). The frequency of acute rejection episodes was lower for the more recent cohort of recipients (9.4% vs 23.9%).

*Survival SL-based CPM*

The survival SL selected only two candidate learners, the Cox model and the RSF, with respective weights of 0.857 and 0.143. Therefore, the SL's predictions were the inverse-logit of the weighted sum of the predictions from the Cox and random survival forest models. Weights were similar when using the binomial logistic likelihood loss and with *a priori* elasticnet predictor screening. However, using the AUROCt loss led to larger weights for random survival forest and survival neural network; see Supplementary Table 2.

*Predictive performance*

Table 2 presents the predictive performance. In DIVAT, the survival SL achieved better discrimination than the KTFS, with a tAUROC of 0.83 (95% CI, 0.79-0.87) compared to 0.76 (CI, 0.70-0.82). Their mean calibration was similar: 0.93 (CI, 0.66-1.26) for the KTFS and 0.92 (CI, 0.78-1.05) for the survival SL. However, for the other calibration measures, the KTFS was better than the survival SL (Figure 3). Regarding the overall performance, the two CPMs were equivalent with a Brier score of 0.10 (CIs, 0.09-0.12 and 0.08-0.11).

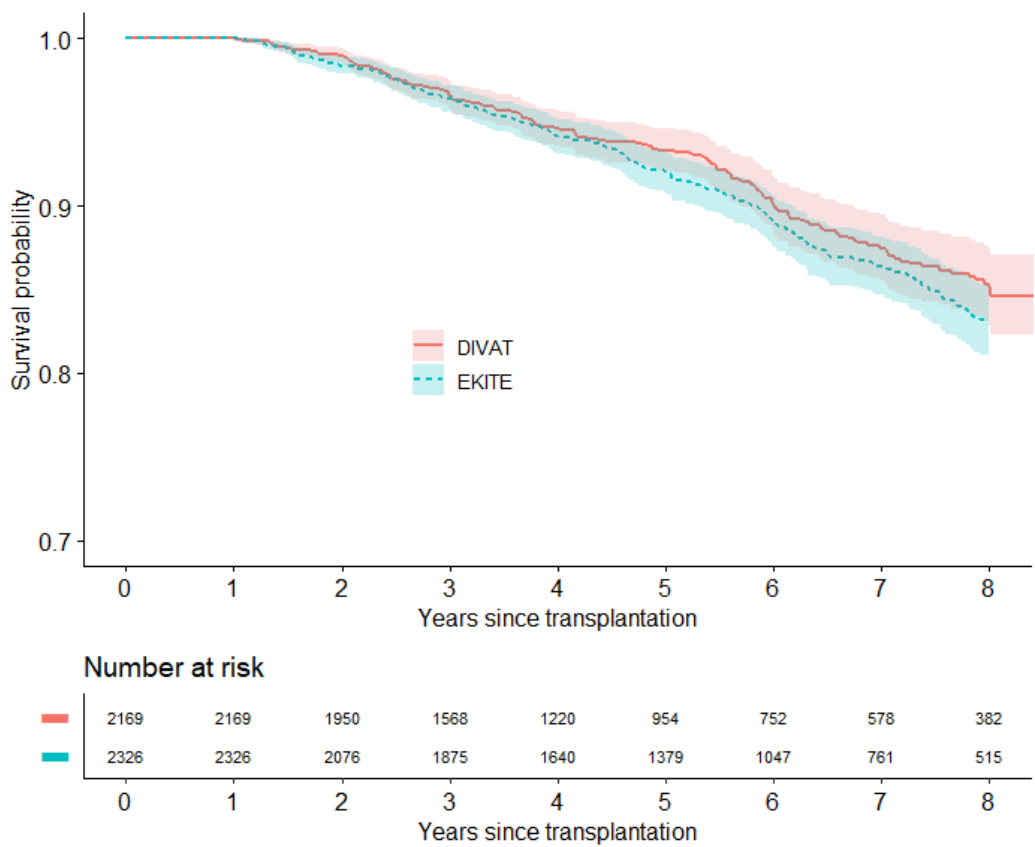

*Figure 2: Survival curves for each cohort. Notes: The survival probability axis is trimmed before 70% to enhance visibility. The samples are selected among the survivors at one year, i.e., the time of the Kidney Transplant Failure Score computation.*

Discrimination was stable over time with a tAUROC of 0.80 (CI, 0.77-0.84) and 0.80 (CI, 0.76-0.83) for the KTFS and the survival SL, respectively. Calibration fluctuated over time for both CPMs, without a clear difference between the two CPMS. Overall performances remained stable over time, with a Brier score of 0.11 (CI, 0.10-0.12) for both KTFS and SL.

*Table 1: Characteristics of the patients for the development in DIVAT and the temporal validation approximately ten years later in EKiTE.*

|  | DIVAT N = 2,169 | EKiTE n = 2,326 | P-value[a] |
|---|---|---|---|
| Recipient age (years), Mean (SD) | 48.0 (13.0) | 53.5 (13.5) | <0.001 |
| Donor age (years), Mean (SD) | 45.2 (15.8) | 55.0 (16.5) | <0.001 |
| Body mass index (kg/m$^2$), Mean (SD) | 23.6 (4.3) | 24.9 (4.4) | <0.001 |
| Cold ischemia time (hours), Mean (SD) | 23.5 (8.8) | 18.3 (6.7) | <0.001 |
| 1-year eGFR (mL/min), Mean (SD) | 51.2 (18.0) | 56.1 (22.9) | <0.001 |
| HLA-incompatibilities, Median [IQR] | 3.0 [2.0 - 4.0] | 3.0 [2.0 - 4.0] | <0.001 |
| Last donor Cr. (µmol/L), Median [IQR] | 86.0 [67.0 - 112.0] | 75.0 [57.0 - 102.0] | <0.001 |
| 3-month Cr. (µmol/L), Median [IQR] | 133.0 [106.0 - 165.0] | 134.0 [108.0 - 168.0] | 0.052 |
| 6-month Cr. (µmol/L), Median [IQR] | 130.0 [106.0 - 160.5] | 133.0 [106.0 - 167.0] | 0.015 |
| 1-year Cr. (µmol/L), Median [IQR] | 130.0 [106.0 - 160.0] | 132.0 [106.0 - 165.0] | 0.197 |
| 3-month Pr. (g/day), Median [IQR] | 0.2 [0.1 - 0.4] | 0.0 [0.0 - 0.0] | <0.001 |
| 6-month Pr. (g/day), Median [IQR] | 0.2 [0.1 - 0.4] | 0.0 [0.0 - 0.0] | <0.001 |
| 1-year Pr. (g/day), Median [IQR] | 0.2 [0.1 - 0.4] | 0.0 [0.0 - 0.0] | <0.001 |
| Male recipients, n (%) | 1,342 (61.9%) | 1,432 (61.5%) | 0.814 |
| Male donors, n (%) | 1,367 (63.3%) | 1,296 (55.7%) | <0.001 |
| Previous transplantation, n (%) |  |  | 0.250 |
| 0 | 1,754 (80.9%) | 1,837 (78.9%) |  |
| 1 | 344 (15.9%) | 407 (17.5%) |  |
| 2+ | 71 (3.3%) | 85 (3.6%) |  |
| Acute rejection in the first year post-transplantation, n (%) | 518 (23.9%) | 218 (9.4%) | <0.001 |
| Abbreviations: Cr., creatinine; eGFR, estimated graft filtration rate; HLA, human leukocyte antigen; IQR, interquartile range; Pr., proteinuria; SD, standard deviation. [a] Continuous variables approximately normally distributed: t-test; Continuous variables not normally distributed: Mann-Whitney test; and categorical variables: Chis-square test. |||||

*Table 2: Performance of the models at the development in DIVAT and the temporal validation approximately ten years later in EKiTE.*

|  | KTFS | | Survival SL | |
|---|---|---|---|---|
|  | DIVAT N = 2,169 | EKiTE N = 2,326 | DIVAT N = 2,169 | EKiTE N = 2,326 |
| **Discrimination** | | | | |
| AUROCt | 0.76 (0.72-0.81) | 0.80 (0.77-0.84) | 0.83 (0.79-0.87) | 0.80 (0.76-0.83) |
| **Calibration** | | | | |
| Mean calibration | 0.93 (0.80-1.08) | 1.16 (1.02-1.30) | 0.92 (0.78-1.05) | 1.06 (0.94-1.20) |
| Weak calibration | 0.94 (0.78-1.06) | 0.83 (0.73-0.94) | 1.32 (1.17-1.45) | 1.14 (1.02-1.27) |
| ICI | 0.01 (0.00-0.04) | 0.03 (0.02-0.05) | 0.04 (0.02-0.05) | 0.03 (0.02-0.05) |
| **Overall performance** | | | | |
| Brier score | 0.10 (0.09-0.12) | 0.11 (0.10-0.12) | 0.10 (0.08-0.11) | 0.11 (0.10-0.12) |

Values presented: Estimated value (95% Confidence interval).
Abbreviations: AUROCt, area under the time-dependent receiver operating curve; ICI, integrated calibration index; KTFS, Kidney Transplant Failure Score.

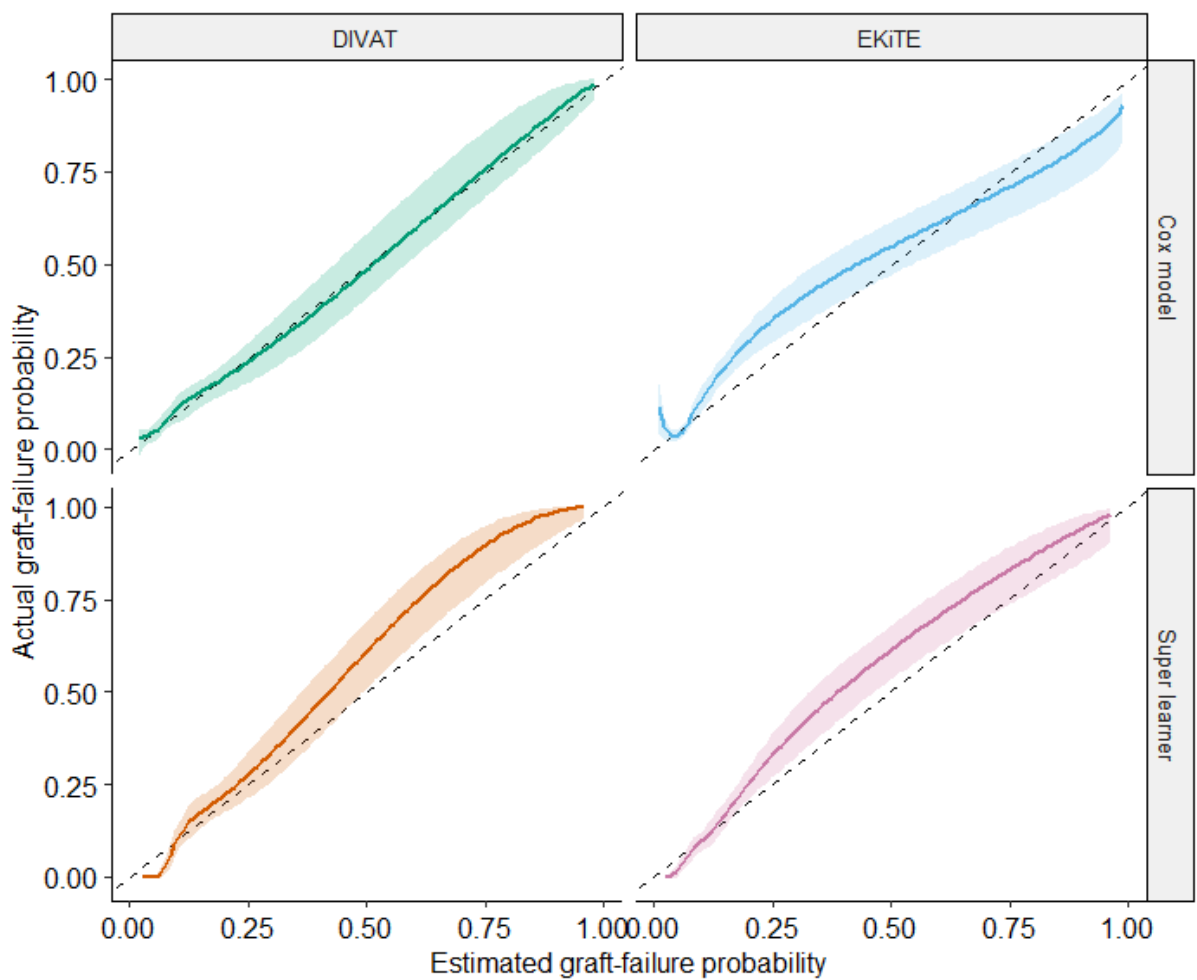

*Figure 3: Flexible calibration curves. Filled areas show the 95% confidence interval.*

Similar trends were observed for the alternative survival SL implementations (Supplementary Table 3). Mean calibration, moderate calibration, and Brier scores were similar to Table 2 and remained stable over time. All other implementations showed worse weak calibrations (including elasticnet screening). As expected, the discrimination was the highest in DIVAT using both AUROCt as loss and with elasticnet screening, reaching 0.94 (CI, 0.92-0.96). However, the AUROCt was similar to the Brier Score-based SL in EKiTE (0.81; CI, 0.77-0.85).

**Discussion**

One of the many challenges researchers face when developing a CPM is choosing a statistical method. Tradeoffs between different approaches may include interpretability, ease of implementation, or statistical performance. In this study, we focused on the stability of performance over time. A steadier CPM is less sensitive to population shift and requires less frequent refitting or updating.[28] We compared a Cox model to a survival SL that combines statistical models and machine learning algorithms. We showed that performance evolution was similar between the two. However, survival SL achieved better discrimination on the population used to fit it at the cost of inferior weak calibration. The survival SL overestimated the graft survival on the population used to fit it, while the KTFS underestimated the graft survival when applied to a future dataset. The calibration slope showed a tendency to decrease over time, perhaps explaining why the survival SL was no longer overestimating the performance in the temporal validation.

Such CPMs' calibration drift occurs because healthcare is a dynamic environment. New patient profiles can be brought by new facilities, giving easier access to healthcare, and new clinical guidelines. The natural ageing of the population may also shift the population characteristics. Alternatively, changing how a predictor is measured can also yield calibration drift.[29] In our motivating case, the donors became older due to the increased use of marginal graft, while the acute rejection prevalence was recently lowered by using higher immunosuppression.[17,30]

Our results differ from those of Davis et al.'s simulation study,[18] which showed that neural networks and random forests yielded better robustness to calibration drift than regression. They used a much larger sample size (n=170,675 during the development period and n=1,671,276 during the validation period), potentially favoring data-hungry machine learning algorithms.[31] The need for larger sample sizes may be further exacerbated by right censoring in our study. Furthermore, methods themselves may deal differently with right censoring, impacting the performance. For instance, Baralou et al. showed through simulations that random survival forests are more sensitive to the censoring level than a Cox proportional hazard model.[32] Lastly, Davis et al. studied regression models with main terms only, while KTFS includes non-linear functional forms and clinically relevant interactions.[4,18] In contrast, our SL prediction was mainly based on a Cox model conditioning on main terms only and with no interactions. Including domain knowledge, such as plausible interactions or relevant predictors, in the SL's candidate learners should likely improve its performance.

Some limitations must be acknowledged. We only investigated one CPM at two time-points, and some of the observed results may be attributable to sampling variability. Benchmarking several CPMs at several times and on several datasets would be ideal. However, many recent literature reviews conclude that CPMs are generally poorly developed,[33–37] challenging the benchmarking of several CPMs. Simulation studies could be another solution to compare machine learning-based and regression-based CPMs, but they are farther from reality and computationally more intensive, while ground truth is not needed to validate prediction.[38] In addition, we did not consider mortality as a competing event but as a censoring event. Therefore, the predicted risk is the hypothetical risk in a counterfactual world in which the competing event does not exist,[39] changing the decision-making process. However, (i) our study is methodological by design and will not impact clinical decision-making, and (ii) only 3.5% of the individuals used at development died during the follow-up. A recent external validation study of the KTFS showed no difference in performance when accounting for competing risk or not.[5]

**Conclusion**
Should we favor regression or machine learning methods to build a CPM? The present study shows that the survival SL (a machine learning technique that may also incorporate regressions) achieved better discrimination and worse weak calibration than a Cox model. In the temporal validation conducted ten years later, both methods were subject to calibration drift. Therefore, population shifts over time did not seem to impact one method more than the other. However, poor calibration at development is often detrimental and temporal recalibration of survival SL risks to maintain the suboptimal weak calibration. In the end, the choice will depend on the CPM's intended use. For instance, one may want to optimize discrimination when the goal is scarce resource allocation (e.g., by prioritizing medically urgent recipient candidates with a high risk of dying without transplantation). However, achieving correct calibration is crucial for patient information and shared decision-making according to the estimated risk of events.

**Data Statement**
The EKiTE network restricts access to clinical data, as these are confidential and are subject to the General Data Protection Regulation. See Lorent et al.[23] for information on requesting access to the data from each EKiTE center's scientific and ethics committee.

**Acknowledgments**
We thank the people involved in the EKiTE network. This study was approved by the Université de Montréal 's clinical ethics committee (#2023-4811). AC was supported by an IVADO postdoctoral fellowship #2022-7820036733. HC is a Fonds de recherche du Québec senior scholar. RWP holds the Albert Boehringer I Chair. MES holds the Canada Research Chair in Causal Inference and Machine Learning in Health Science.

**Author Contributions**
AC designed the study. AC and KAF were involved in data collection and curation. EP performed the analyses with input from AC, MES and RWP. All authors interpreted the results. AC wrote the manuscript, which was revised by all coauthors.

**Supplementary materials of "What if we had built a prediction model with a survival super learner instead of a Cox model 10 years ago?" from Chatton, Pilote, Assob Feugo, Cardinal, Platt & Schnitzer**

Supplementary table 1: Reproducibility information.

| Purpose | R Package | Version | Authors | Learner or measure | Hyperparameters |
|---|---|---|---|---|---|
| Candidate learners | flexsurv [1] | 2.2.1 | Jackson C. | Gamma-distributed AFT model | - |
| | | | | Weibull-distributed AFT model | - |
| | | | | Royston & Parmar model [2] | k = 3 |
| | survival | 3.4-0 | Therneau T. | Cox model | - |
| | glmnet [3,4] | 4.1-6 | Friedman J., Hastie T., Tibshirani R., Narasimhan B., Tay K., Simon N., Qian J. & Yang J. | Elasticnet Cox model | α = 0.9 (0.2[†]) |
| | | | | | λ = 0.003 (0.028[†]) |
| | randomForestSRC | 3.2.0 | Ishwaran H. & Kogalur U.B. | Random survival forest [5] | nodesize = 20 |
| | | | | | mtry = 3 |
| | | | | | ntree = 500 |
| | survivalmodels | 0.1.13 | Sonabend R. | Survival neural network [6] | n.nodes = 20 |
| | | | | | decay = 0.1 |
| | | | | | batch.size = 256 |
| | | | | | epochs = 1 |
| Python to R conversion | reticulate | 1.28 | Ushey K., Allaire J. & Tang Y. | - | - |
| Super Learner | survivalSL | 0.94 | Foucher Y. & Sabathe C. | - | - |
| Performance measures | timeROC | 0.4 | Blanche P. | AUROCt [7] | - |
| | rms | 6.4-1 | Harrel Jr F.E. | Calibration | - |
| | riskRegression | 2022.11.28 | Gerds T.A. & Kattan W. (8) | Brier Score | - |

R version: 4.2.2, Python version: 3.12 (Windows 10 x64), seed set randomly to 9511.
Abbreviations: AFT, Accelerated Failure Time; AUROCt, Area Under the time-dependent Receiver Operating Curve.
[1] Jackson C. (2016). Journal of Statistical Software, 70(8), 1-33. [2] Royston P. and Parmar M. (2002). Statistics in Medicine 21(1):2175-2197. [3] Friedman J., Hastie T. and Tibshirani R. (2010). Journal of Statistical Software, 33(1), 1-22. [4] Simon N., Friedman J., Hastie T. and Tibshirani R. (2011). Journal of Statistical Software, 39(5), 1-13. [5] Ishwaran H., Kogalur U.B., Blackstone E.H. and Lauer M.S. (2008). Ann. Appl. Statist. 2(3), 841–860. [6] Katzman J. L., Shaham U., Cloninger A., Bates J., Jiang T., and Kluger, Y. (2018). BMC Medical Research Methodology, 18(1), 24. [7] Blanche P., Dartigues J.-F. and Jacqmin-Gadda H. (2013). Statistics in Medicine, 32(30), 5381-5397. [8] Gerds T.A. and Kattan M.W. (2021). Medical Risk Prediction Models: With Ties to Machine Learning (1st ed.) Chapman and Hall/CRC.
[†] Value for screening.

Supplementary Table 2: Candidate learners' weights by survival super learner

implementations.

|  | All KTFS predictors | | | Elasticnet screening | | |
| --- | --- | --- | --- | --- | --- | --- |
| Candidate learner | BS | NBLL | AUROC | BS | NBLL | AUROC |
| Gamma-distributed AFT model | 0 | 0.02 | 0 | 0 | 0 | 0 |
| Weibull-distributed AFT model | 0 | 0 | 0.15 | 0 | 0 | 0.03 |
| Cox model | 0.86 | 0.63 | 0.14 | 0.61 | 0.74 | 0.17 |
| Elasticnet Cox model [†] | 0 | 0 | 0.10 | - | - | - |
| Royston-Parmar model | 0 | 0 | 0.04 | 0 | 0 | 0.02 |
| RSF | 0.14 | 0.35 | 0.49 | 0.39 | 0.26 | 0.71 |
| SNN | 0 | 0 | 0.07 | 0 | 0 | 0.06 |

Abbreviations: AFT, Accelerated Failure Time; AUROC, Area Under The Receiver Operating Curve; BS, Brier Score; NBLL, Negative Binomial log-Likelihood; RSF, Random Survival Forest; and SNN, Survival Neural Network.
[†] Elasticnet was removed from the pool of candidate learners when used for screening.

Supplementary Table 3: *Performance of the alternative survival super learners at the development and approximately ten years later.*

| | All KTFS predictors | | | | Elasticnet screening | | | | | |
|---|---|---|---|---|---|---|---|---|---|---|
| | NBLL | | AUROCt | | Brier Score | | NBLL | | AUROCt | |
| | DIVAT N = 2,169 | EKiTE N = 2,326 | DIVAT N = 2,169 | EKiTE N = 2,326 | DIVAT N = 2,169 | EKiTE N = 2,326 | DIVAT N = 2,169 | EKiTE N = 2,326 | DIVAT N = 2,169 | EKiTE N = 2,326 |
| **Discrimination** | | | | | | | | | | |
| AUROCt | 0.87 (0.84-0.90) | 0.78 (0.74-0.82) | 0.93 (0.91-0.95) | 0.82 (0.78-0.85) | 0.88 (0.85-0.91) | 0.81 (0.78-0.85) | 0.84 (0.80-0.88) | 0.81 (0.78-0.85) | 0.94 (0.92-0.96) | 0.81 (0.77-0.85) |
| **Calibration** | | | | | | | | | | |
| Mean calibration | 0.92 (0.78-1.06) | 1.08 (0.95-1.22) | 0.97 (0.82-1.10) | 1.06 (0.94-1.18) | 0.94 (0.80-1.08) | 1.16 (1.01-1.32) | 0.92 (0.78-1.06) | 1.16 (1.00-1.31) | 0.95 (0.82-1.09) | 1.11 (0.96-1.25) |
| Weak calibration | 1.51 (1.34-1.66) | 0.91 (0.76-1.09) | 2.09 (1.90-2.28) | 1.57 (1.39-1.76) | 1.55 (1.36-1.72) | 1.24 (1.09-1.41) | 1.39 (1.22-1.55) | 1.26 (1.09-1.43) | 2.13 (1.86-2.35) | 1.34 (1.16-1.54) |
| ICI | 0.06 (0.04-0.07) | 0.03 (0.02-0.05) | 0.09 (0.07-0.11) | 0.05 (0.03-0.06) | 0.05 (0.03-0.06) | 0.03 (0.02-0.05) | 0.04 (0.02-0.05) | 0.03 (0.02-0.05) | 0.08 (0.06-0.10) | 0.03 (0.02-0.05) |
| **Overall performance** | | | | | | | | | | |
| Brier score | 0.09 (0.08-0.11) | 0.11 (0.10-0.12) | 0.09 (0.08-0.10) | 0.11 (0.10-0.12) | 0.09 (0.08-0.11) | 0.10 (0.09-0.12) | 0.10 (0.08-0.11) | 0.10 (0.09-0.12) | 0.08 (0.07-0.09) | 0.10 (0.09-0.12) |

Values presented: Estimated value (95% Confidence interval).
Abbreviations: AUROCt, area under the time-dependent Receiver Operating Curve; ICI, integrated calibration index; KTFS, Kidney Transplant Failure Score; NBLL, Negative Binomial log-Likelihood.